\documentclass[12pt]{iopart}

\usepackage{graphicx}
\usepackage{tabularx}
\usepackage{comment}
\usepackage[dvipsnames]{xcolor}
\usepackage{amsmath,amssymb}
\usepackage{mathrsfs}
\usepackage{hyperref}
\begin{document}

\title{Defect - related Anomalous Mobility of Small Polarons in Oxides: the Case of Congruent Lithium Niobate}

\author{Anton Pfannstiel$^1$, Mirco Imlau$^1$, Marco Bazzan$^2$ and Laura Vittadello$^1$}

\address{$^{1}$ Institute of Physics, Department of Mathematics/Informatics/Physics, University of Osnabr\"uck, Barbarastra\ss{}e 7, D-49076 Osnabr\"uck, Germany}
\address{$^{2}$ Università di Padova, Dipartimento di Fisica e Astronomia, Via Marzolo 8, 35131, Padova, Italy
}
\ead{laura.vittadello@uni-osnabrueck.de}
\vspace{10pt}
\begin{indented}
\item[]April 2024
\end{indented}

\begin{abstract}
Polarons play a major role in the description of optical, electrical and dielectrical properties of several ferroelectric oxides.
The motion of those particles occur by elementary hops among the material lattice sites. In order to compute macroscopic transport parameters such as charge mobility, normal (i.e. Fickian) diffusion laws are generally assumed.
In this paper we show that when defect states able to trap the polarons for long times are considered, significant deviations from the normal diffusion behaviour arise.
As an example of this behavior, we consider here the case of lithium niobate (LN). This can be considered as a prototypical system, having a rich landscape of interacting polaron types and for which a significant wealth of information is available in literature. 
Our analysis considers the case of a stoichiometric, defect-free lithium niobate containing a certain concentration of small electron polarons hopping on regular Nb sites, and compares it to the material in congruent composition, which is generally found in real-life applications and which is characterized by a large concentration of antisite $\mathrm{Nb_{Li}}$ defects. While in the first case the charge carriers are free polarons hopping on a regular Nb sublattice, in the second case a fraction of polarons is trapped on  antisite defects. Thus, in the congruent material, a range of different hopping possibilities arises, depending on the type of starting and destination sites. 
We develop a formalism encompassing all these microscopic processes in the framework of a switching diffusion model which can be well approximated by a mobile-immobile transport model  providing explicit expressions for the polaron mobility. 
Finally, starting from the Marcus-Holstein's model for the polaron hopping frequency we verify by means of a Monte Carlo approach the diffusion/mobility of the different polarons species showing that, while free polarons obeys the laws for normal diffusion as expected, bound polarons follow an anomalous diffusion behaviour and that in the case of the congruent crystal where mixed free and bound polaron transport is involved, our expressions indeed provide a satisfactory description. 
\end{abstract}

\section{Introduction}

In the last decades it became apparent that the charge transport in a number of polar materials must be understood in terms of polaron motion. These are quasi-particles made up of an electrical charge that, by interaction with the crystalline environment, is able to distort the local lattice creating a local potential well: as a net result, the particle becomes self-localized \cite{Franchini2021}. Polarons play a major role when it comes to the interpretation of the optical, electrical or dielectrical properties \cite{Maglione2016} of important technological materials such as $\mathrm{LiNbO_3}$ \cite{Schirmer2009,Vittadello2018}, $\mathrm{KNbO_3}$ \cite{Grigorjeva2004,Torbruegge2008}, $\mathrm{LiTaO_3}$ \cite{Kappers1985, Liu2004}, $\mathrm{BaTiO_3}$ \cite{Tsunoda2019, Possenriede1994}, $\mathrm{PbTiO_3}$ \cite{Ghorbani2022}, $\mathrm{r-TiO_2}$ \cite{Bogomolov2022, Tanner2021}, $\mathrm{CeO_2}$ \cite{Tuller1977, Plata2013} and solid solutions of the above like $\mathrm{LiNb_{1-x}Ta_xO_3}$ \cite{Doemer2024}  or $\mathrm{KTa_{1-x}Nb_xO_3}$ \cite{Liu1991}. 
In these polar materials, the electron-phonon coupling is so strong that the exceeding charge is localized on a single lattice site, i.e. only \emph{small} strong - coupling polarons are formed. These polarons can move in the lattice via random hopping mechanisms in response to the lattice thermal motion, as described by the Marcus-Holstein theory \cite{Holstein1959}. 
In standard formulations used to compute the drift mobility of the polaronic carriers, it is assumed that the latter hop on a regular lattice of spacing $d$ among by nearest-neighboring sites. Thus the mobility of the polarons follows the familiar law  $\mu \propto ed^{2}w/(k_\mathrm{B}T)$ \cite{Austin2001, Alexandrov2007}.

However, the Marcus- Holstein model forecasts an exponential distance-dependent hopping frequency: depending on the characteristic hopping length, the hopping process may involve also far away hopping sites at a distance bigger than $d$. Even more daunting is the fact that real polar materials are often characterized by the presence in the lattice of a considerable amount of point defects. Due to extra Coulomb interaction, they can act as privileged localization centers for the polarons. When sitting on those sites, the hopping frequency at a given temperature may differ of several orders of magnitude with respect to free polarons moving on regular sites. Thus different types of polarons may exist in a given material and
their interplay can result in a rich and complex behaviour, like the formation of polarons bound to defects (bound polarons), polaron complexes made up of a regular and a bound polaron (bipolarons) \cite{Koppitz1987, Schmidt2020} or a hole polaron and an electron polaron (self-trapped excitons) \cite{Messerschmidt2018}. 
In recent works \cite{Mhaouech2016, Vittadello2018, Guilbert2018, Vittadello2021} it has been shown that the relative weight of the different microscopic hopping processes is strongly dependent on the temperature and on the defect concentration.

The goal of the present work is to provide a novel formulation for the calculation of the polaron mobility taking into account the complications induced by the presence of defects and of the different types of polarons arising from them.
As a case study, we will consider $\mathrm{LiNbO_3}$, (LN). This is a technologically important material widely used in a great variety of applications ranging from electro-optic modulators, waveguides \cite{Vittadello2016}, nonlinear optical devices, photorefractive holography and, more recently, for novel applications such as the manipulation of small organic \cite{BlazquezCastro2011} and inorganic objects \cite{Lucchetti2016}, as substrate for photorefractive tweezers \cite{Villarroel2011} and as a platform for microfluidic chips \cite{wu2019acoustofluidic, Zanini2022, cremaschini2024trifurcated, Zamboni2021}.
The structure of stoichiometric LN is constituted by a quasi-cubic Nb sublattice intercalated by Li and O atoms. In an ideal defect-free material (stoichiometric LN, sLN) polarons can only form at the regular Nb sites $\mathrm{Nb}_\mathrm{Nb}^{5+}$ ions  and are indicated as \emph{F} polarons. 
However, LN is normally produced at the congruent composition (cLN) which is characterized by a high concentration (typically, up to 1 mol.\%) of  substitutional ``antisite'' defects $\mathrm{Nb_{Li}^{5+}}$ \cite{Volk2008}. These defects constitute the preferential sites for the formation of polarons indicated as bound polarons (\emph{B} polarons). While further polaron types and localization sites exist, they contribute to a lesser extent to the transport of electrons and are thus omitted in this work.
In LN the relevant microscopic parameters necessary to compute the Marcus-Holstein hopping frequency are known from previous studies \cite{Schirmer2009, Guilbert2018, Vittadello2018}, making it an excellent test case. Our approach can anyway be adapted to different materials.

\section{Theory}

Polarons move through the crystal via hopping transport. According to the Marcus-Holstein theory, each transition is attributed a hopping frequency of the following type: \cite{Guilbert2018, Holstein1959}:
\begin{subequations}
    \begin{equation}
         w_{if}(r,T) = w_{if}^0(T)\exp \left(-\frac{r_{if}}{a_{if}} \right)
         \label{eq:hopping_freq}
    \end{equation}
    \begin{equation}
        w_{if}^0(T) = \frac{1}{2}\left(\frac{\pi}{k_\mathrm {B}T\lambda_{if}} \right)^{\frac{1}{2}} \frac{I_{if}^{2}}{\hbar} \exp \left(-\frac{U_{if}}{k_\mathrm{B}T} \right)
        \label{eq:MH_trap_prefac}
    \end{equation}
    \begin{equation}
    U_{if}= \frac{ (2E_i-\varepsilon_i+\varepsilon_f - e( \textbf{r}_i-\textbf{r}_f) \cdot \textbf{F})^2}{4(E_i+E_f)}
    \end{equation}
    \label{eq:hopping_barr}
\end{subequations}
where $k_\mathrm{B}$ is the Boltzmann constant, $T$ the absolute temperature, $\hbar$ the reduced Planck constant and $e$ the elementary charge. The lattice reorganization energy $\lambda_{if} = E_i + E_f$ is spent to rearrange the lattice upon hopping, $I_{if}$ is the hopping integral pre-factor, describing the electronic wave-function overlap between neighboring sites, $a_{if}$ is the orbital parameter describing the localization strength of the electronic wave-function, $U_{if}$ is the energy barrier for the hopping process, $\varepsilon_i$ is the pre-localization energy of the electron at zero deformation, $E_i$ is the polaron stabilization energy, $\mathbf{F}$ is an applied electrical field and $\textbf{r}_{i}$, $\textbf{r}_{f}$ the position vector of the initial or final site respectively. 

\subsection{Free polaron hopping}
\label{sec:ff_theory}

 As a first case, we consider undoped, defect free stoichiometric LN. In this case excess charge carriers introduced in the lattice e.g. by thermal reduction treatments, form free polarons hopping randomly within the ordered structure of the $\mathrm{Nb_{Nb}}$ sublattice. In these conditions, all the sites are equivalent and the distribution of the waiting times between the jumps
satisfies the hypothesis of the Central Limit Theorem, which lead to the normal diffusion behavior \cite{Metzler2000}. Macroscopically, this results in the standard Fick's law. In a 1-dimensional case:
 
\begin{equation}
\frac{\partial P(z,t)}{\partial t}=D_\mathrm{FF}\frac{\partial^{2}P(z,t)}{\partial z^{2}}
\label{eq:Fick's law}
\end{equation}
where $D_\mathrm{FF}$ is the diffusion coefficient for free polarons and $P(z,t)$ is the Particle Distribution Function (PDF). 
 For free polarons, the diffusion coefficient can be calculated in the following form \cite{Metzger1984}:
 \begin{equation}
    D_\mathrm{FF}  = \frac{1}{6}\sum_{j=1}^{N}r_{j}^{2}w_\mathrm{FF}(r_{j},T) = \frac{w_\mathrm{FF}^0(T)}{6}\sum_{j=1}^{N}r_{j}^{2}\exp{\left(-\frac{r_{j}}{a_\mathrm{FF}}\right)}
    \label{eq:diff_teo}
\end{equation}
where the summation runs over all the sites of the lattice, $r_j$ is the distance from the origin to the $j$-th neighbour, and $w_\mathrm{FF}$ is the hopping frequency of equation \ref{eq:hopping_freq}. Even though the exponential decay of the Holstein's frequency guarantees that the summation converges quite fast, in order to get accurate results it is necessary to consider beyond-nearest neghbouts contributions. 
For practical purposes, the sum \ref{eq:diff_teo} it can be truncated up to a certain coordination sphere around the starting site \cite{Vittadello2018}. 
The solution to eq. \ref{eq:diff_teo} at time $t$ for an initial delta-like distribution placed at the origin $P(z,0)=\delta(z)$ is the Gaussian distribution:

\begin{equation}
    P(z,t) = \frac{1}{\sqrt{4 \pi D_\mathrm{FF} t}} \exp{ \left( -\frac{z^{2}}{4 D_\mathrm{FF} t} \right)}
    \label{eq: Gaussian}
\end{equation}
As it is well known, this distribution has zero mean and variance increasing linearly with time of the form
\begin{equation}
\sigma^{2}_{\rm FF} = 2 D_\mathrm{FF} t
\end{equation}

If an electric field is present, the average of the PDF remains no longer at zero, but instead increases linearly with time with a given mobility $\mu$:

\begin{equation}
<z>= \mu_\mathrm{FF} \mathrm{F} t.
\label{eq:mean_free}
\end{equation}
with $\mathrm{F}=|\bf{F}|$ the electric field strength, which here is assumed oriented along $z$.

The knowledge of the diffusion coefficient is sufficient to describe completely the mobility via the Einstein's relationship $\mu_\mathrm{FF}=eD_\mathrm{FF}/(k_\mathrm{B}T)$ as long as the stochastic behaviour dominates the electric field one \cite{Nenashev2010}. This condition is satisfied when $e\bf{r \cdot F}$ $\ll k_\mathrm{B} T$.

\subsection{Bound polaron hopping}
\label{sec:pp_theory}
Congruent lithium niobate is the most commonly encountered composition of this material, since it is the energetically most stable and the one that is most easily grown. It contains about  $[\mathrm{Nb_{Li}}] = 19.09 \cdot 10^{25}\,\mathrm{m^{-3}}$ antisite defects \cite{Volk2008}. 
Those antisites are localization centers for B polaron formation.  
As it was reported in several recent papers \cite{Vittadello2018, Guilbert2018, Vittadello2021}, for sufficiently low temperatures and congruent composition, the main channel for polaron transport is direct hopping among the antisites which can be viewed collectively as the sites of a disordered lattice. 
Due to this increased disorder level and to the exponential dependence of the hopping frequency, the Central Limit Theorem is no longer applicable to the distribution of the waiting times and the normal diffusion theory breaks down. The theoretical framework to deal with this problem is the Continuous Time Random Walk (CTRW) model \cite{Metzler2000}  in which a particle, after hopping to a given site, waits a certain time before moving again. 
For this situation, the Fick's second law (Eq. \ref{eq:Fick's law}) is replaced by a more general equation governing the diffusion. In 1-D, it can be expressed in the following form \cite{Metzler2000}:

\begin{equation}
\frac{\partial P(z,t)}{\partial t}={_0}D_{t}^{1-\alpha}D_\mathrm{BB}\frac{\partial^{2}P(z,t)}{\partial z^{2}}
\label{eq:anomal_equation}
\end{equation}
where P(z,t) is the PDF, ${_0}D_{t}^{1-\alpha}$ is the fractional derivative operator defined in Ref. \cite{Metzler2000}, $D_{\rm BB}$ is the generalized diffusion coefficient for bound polarons, having the dimension $[D_{\rm BB}]=\frac{m^{2}}{s^{\alpha}}$ and $\alpha$ is the anomalous parameter classifying the type of diffusion. 
For $\alpha=$1, Eq. \eqref{eq:anomal_equation} becomes the standard Fick's equation so that this case describes \emph{normal} diffusion while all other cases are termed \emph{anomalous}. 
Thus we may expect that, in contrast to the free polaron case occurring in defect-free materials, bound polarons obey eq. \ref{eq:anomal_equation} if the proper conditions are met. In the case of bound polarons diffusing in LN, as it will be shown below, it is found $\alpha<1$, which corresponds to the so-called sub-diffusive case.
The most general fundamental solution of equation \eqref{eq:anomal_equation} for the sub-diffusive case is given in terms of the Fox H-function \cite{Montroll1965} which can be expressed via the following series expansion:
 
\begin{equation}
P(z,t)=\frac{1}{\sqrt{4D_{\rm BB}t^{\alpha}}}\sum_{n=0}^{\infty}\frac{(-1)^{n}}{n!\Gamma(1-\alpha[n+1]/2)}\left(\frac{z^{2}}{D_{\rm BB}t^{\alpha}}\right)^{n/2}
\label{eq:Fox_expansion}
\end{equation}

The 1D variance of the distribution of equation \eqref{eq:Fox_expansion} has the form of:

 \begin{equation}
 \sigma^{2}_{\rm BB}(t) =\frac{2D_{\rm BB}t^{\alpha}}{\Gamma(1+\alpha)}
\label{eq:sigma_bound}
\end{equation}

and it is no longer proportional to the first power of time. 

If the stochastic behaviour dominates the electric field one, it can be shown that the \emph{anomalous} mobility can still be computed from the generalized diffusion coefficient via the Einstein's equation:

 \begin{equation}
 \mu_{\alpha}= \frac{eD_{\rm BB}}{k_BT}
 \end{equation}
 The first moment of the distribution is now described by the relationship:

 \begin{equation}
 \left\langle z(t)\right\rangle _{F}=\frac{F\mu_{\rm BB}t^{\alpha}}{\Gamma(1+\alpha)}
\label{mean_value_bound}
\end{equation}

\subsection{Free and bound mixed hopping }
\label{sec:congr_theory}
The most general situation, which is also the one most often encountered in experiments at room temperature with congruent LN is the one where both bound and free polarons contribute to the transport. 
In this case two more elementary hopping processes besides FF and BB hopping are involved in the transport, i.e. BF and FB. The idea here is to relate the two transport modes detailed in sections \ref{sec:ff_theory} and \ref{sec:pp_theory} via some coupled transport models.  

\subsubsection{Switching Diffusion model.}
\begin{figure}
\centering
\includegraphics{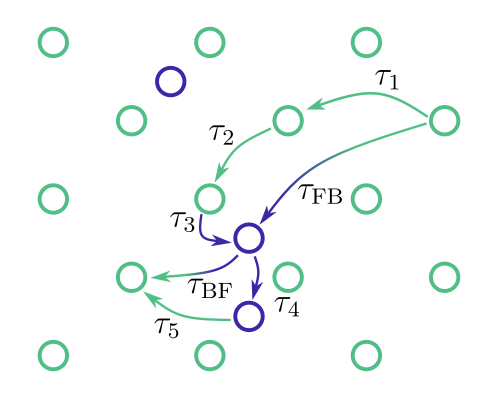}
\caption{Schematic representation of switching time between F (green dots) and B (blue dots) polarons used on the MIM and SW model. In the scheme depicted here $\tau_\mathrm{FB}=\tau_1+\tau_2+\tau_3$ and $\tau_\mathrm{BF}=\tau_4+\tau_5$.}
\label{fig:sketch}
\end{figure} 

In the switching diffusion model, the population of F and B polarons is coupled through some switching rates $k_{ij}$, whose physical meaning is sketched in Fig. \ref{fig:sketch}. They represent the inverse of the average time  between sequential changes of species, irrespective of the path that the polaron takes between the species change: $\tau_{ij} = k_{ij}^{-1}$. In the case studied here, the model described in Ref. \cite{Grebenkov2019} is used as reference and extended to our case to account for the fact that F polarons obey the normal diffusion case, while B follow an anomalous behavior. The coupled diffusion equations then take the form:

\begin{subequations}
\begin{equation}
    \frac{\partial P_{\rm F}(x,t)}{\partial t}=D_{\rm FF}\frac{\partial^2 P_{\rm F}(x,t)}{\partial x^2}-k_{\rm FB}P_{\rm F}(x,t)+k_{\rm BF}P_{\rm B}(x,t)
\end{equation}    
\begin{equation}
    \frac{\partial P_{\rm B}(x,t)}{\partial t}= {}_0D_t^{1-\alpha}D_{\rm BB}\frac{\partial^2 P_{\rm B}(x,t)}{\partial x^2}+k_{\rm FB}P_{\rm F}(x,t)-k_{\rm BF}P_{\rm B}(x,t)
    \label{eq:switching_differential}
\end{equation}
\end{subequations}
with $P_{\rm F}(x,t)$ and $P_{\rm B}(x,t)$ the PDFs of the free and bound polarons and $D_{\rm FF}$ and $D_{\rm BB}$ their respective diffusion coefficient. The Riemann-Liouville fractional derivative operator ${}_0D_t^{1-\alpha}$ in equation \ref{eq:switching_differential} takes into account the anomaly of the bound polaron transport \cite{Metzler2000,Metzler2022}. 

Application of the Laplace transform $\mathscr{L}\{P_i(x,t)\}=\hat{P_i}(x,s)$ simplifies the expressions according to:
\begin{subequations}
\begin{equation}
    s\hat{P_{\rm F}}(x,s)-P_{\rm F}(x,t=0)=D_{\rm FF}\frac{\partial^2 \hat{P_{\rm F}}(x,s)}{\partial x^2}-k_{\rm FB}\hat{P_{\rm F}}(x,s)+k_{\rm BF}\hat{P_{\rm B}}(x,s)
\end{equation}    
\begin{equation}
    s\hat{P_{\rm B}}(x,s)-P_{\rm B}(x,t=0)= s^{1-\alpha}D_{\rm BB}\frac{\partial^2 \hat{P_{\rm B}}(x,s)}{\partial x^2}+k_{\rm FB}\hat{P_{\rm F}}(x,s)-k_{\rm BF}\hat{P_{\rm B}}(x,s)
\end{equation}
\end{subequations}
where we applied $\mathscr{L}\{ {}_0D_t^{1-\alpha}f(x,t)\}=s^{1-\alpha}\hat{f}(x,s)$ \cite{Metzler2000}. This expression is analytically solvable in the Laplace space and a real space solution can be derived by numerical inverse Laplace transform. 
The process is performed separately for the two cases of polarons starting as F or as B polarons respectively. In the former case the boundary conditions are chosen as $P_{\rm F}(x,t=0)=\delta(x)$ and $P_{\rm B}(x,t=0)=0$ and reciprocally $P_{\rm F}(x,t=0)=0$ and $P_{\rm B}(x,t=0)=\delta(x)$ for the latter, with $\delta(x)$ the Dirac delta function. 
Additionally, for all cases $\lim_{b\to \infty} P_{\rm F}(\pm b,t)=\lim_{b\to \infty} P_{\rm B}(\pm b,t)=0$ is chosen as an additional boundary condition. 
It is then possible to have access to both PDFs from which the variance of the distribution can be computed as a function of time as well as the mobility thanks to the Einstein's equation.  

\subsubsection{Mobile - Immobile model.}
The previous treatment allow for an exact numerical solution of the problem, but does not have an analytical solution.
Since the hopping frequency of bound polarons for all the processes considered here is several orders of magnitude smaller than the one of free polarons, we expect that a good approximation of the previous situation may be provided by a  Mobile-Immobile diffusion model (MIM)\cite{Doerries2022} which on the contrary is analytically solvable.
In the MIM model, free polarons are considered the mobile specie, normally diffusing with $D_{\mathrm{FF}} > 0$ and bound polaron are considered immobile with  $D_{\mathrm{BB}} \approx 0$. The mobile species immobilize with a rate $k_{\rm FB}$ and become mobile again with a rate $k_{\rm BF}$.
In this case the diffusion equations take the form:

\begin{subequations}
\begin{equation}
    \frac{\partial P_{\rm F}(x,t)}{\partial t}= D_{\rm FF}\frac{\partial^2 P_{\rm F}(x,t)}{\partial x^2} - k_{\mathrm{FB}} P_{\rm F}(x,t) + k_{\mathrm{BF}} P_{\rm B}(x,t)
\end{equation}

\begin{equation}
    \frac{\partial P_{\rm B}(x,t)}{\partial t}= k_{\mathrm{FB}} P_{\rm F}(x,t) - k_{\mathrm{BF}} P_{\rm B}(x,t) 
\end{equation}
\end{subequations}

In the MIM model, the variances of the distribution can be computed analytically as \cite{Doerries2022}:
\begin{subequations}
    \begin{equation}
        \sigma^2= \frac{2D_{\rm FF}}{1+ k_{\mathrm{FB}}/ k_{\mathrm{BF}}} \left[t+ \frac{k_{\mathrm{FB}}/ k_{\mathrm{BF}}^2}{1+ k_{\mathrm{FB}}/ k_{\mathrm{BF}}}\left( 1- e^{- (k_{\mathrm{BF}}+k_{\mathrm{FB}})t}   \right) \right]
        \label{eq:doerris_msd}
    \end{equation}
    \begin{equation}
        \sigma^2= \frac{2D_{\rm FF}}{1+ k_{\mathrm{FB}}/ k_{\mathrm{BF}}} \left[t- \frac{k_{\mathrm{BF}}^{-1}}{1+ k_{\mathrm{FB}}/ k_{\mathrm{BF}}} \left( 1- e^{- (k_{\mathrm{BF}}+k_{\mathrm{FB}})t}   \right) \right]
        \label{eq:doerris_msd_startbound}
    \end{equation}
\end{subequations}
for free and bound starting site, respectively. 

The displacement of the polaron distribution under the effect of a bias can be readily computed from the Einstein's relation providing:
\begin{equation}
    <z>= \frac{e F \sigma^2}{2k_BT}
    \label{eq:z_avg_doerris}
\end{equation}
where $\sigma$ stems for the variance in one of the two cases of initial F- or B- polaron population. 
Clearly, in this case we shall define a time-dependent \emph{instantaneous} mobility $\mu(t) = \frac{d \langle z \rangle}{d t} $  as the average displacement is no longer proportional to time.

\section{Monte Carlo simulations of polaron hopping transport in LN}
Monte Carlo (MC) simulations have proven themselves a valuable tool in modelling polaron transport in LN \cite{Guilbert2018, Vittadello2018}. A dedicated Monte Carlo algorithm is developed to study the material transport properties using the same methodology detailed in the cited papers.
The code generates a stoichiometric $\mathrm{LiNbO_3}$ structure with periodic boundary conditions and subsequently a certain number of  $\mathrm{Nb_{Li}}$ antisites is randomly placed in the structure in the appropriate lattice sites, according to the crystal composition. 
 
The hopping frequencies for a given initial site $i$ towards a destination $f$ included in a suitable volume around the starting site are computed by eq. \ref{eq:hopping_freq} using the values reported in Tab. \ref{tab:param} \cite{Guilbert2018, Vittadello2018, Schirmer2009}. The final destination is chosen via the Gillespie algorithm.
The time needed to perform the hop is then recorded and the process is repeated until a given time has elapsed. A new polaron is then created and the whole simulation iterated until a satisfactory statistics has been achieved. 
For each run, the program records the final polaron positions with respect to its original site, the number of different sites encountered during its walk and the time needed to reach the final site. All the information is then analyzed to compute the variance and the mean displacement of the final polaron distribution as a function of  time. In the simulation we consider the possible presence of a static electric field $\textbf{F}$ directed along the +\textbf{z} direction to study polaron mobility, while for the study of diffusion processes the field is set at zero. 

\begin{table}[]
    \centering
    \begin{tabular}{c|c|c}
    \hline
     Parameters  & Value & Reference \\
     \hline
     $E_\mathrm{F}$    & 0.545\,eV   & \cite{Schirmer2009}\\ 
     $E_\mathrm{B}$    & 0.750\,eV     &  \cite{Guilbert2018}\\
     $\varepsilon_\mathrm{F}$   & 0.000\,eV & \cite{Schirmer2009}\\
     $\varepsilon_\mathrm{B}$   & 1.690\,eV &\cite{Guilbert2018}\\
     $a_\mathrm{FF}=a_\mathrm{FB}=a_\mathrm{BB}=a_\mathrm{BF}$   & 1.60\,\AA & \cite{Guilbert2018}\\
     $I_\mathrm{FF}=I_\mathrm{FB}=I_\mathrm{BB}=I_\mathrm{BF}$   & 0.30\,\AA & \cite{Guilbert2018}\\
     \hline
    \end{tabular}
    \caption{Parameters used to compute the hopping frequency in Eq. \ref{eq:hopping_freq}}
    \label{tab:param}
\end{table}

\section{Results}
\subsection{Free polaron hopping}
\label{sec:ff_results}

\begin{figure}
\centering
\includegraphics{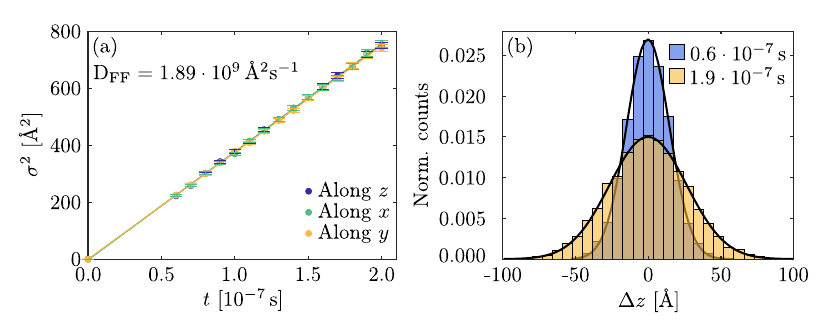}
\caption{(\textbf{a}) Variance of the final electron position as a function of the time on an ordered $\mathrm{Nb_{Nb}}$ structure at $\mathrm{T=300\,K}$  (\textbf{b}) Normalized probability distribution at two different times of the population of F polarons diffusing on the same lattice as in (\textbf{a}) at $\mathrm{T=300\,K}$. Both histograms are fitted with a Gaussian distribution (black curves).}
\label{fig:free}
\end{figure}

As a first case, we consider a system of free polarons hopping in the $\mathrm{Nb_{Nb}}$ sublattice. Fig. \ref{fig:free}a) show the variance of the polaron distribution as a function of the time along the $x$, $y$, and $z$ crystallographic directions in a stoichiometric LN crystal at room temperature. 
As expected, the variance increases linearly with time. This is the fingerprint that the polaron is normally diffusing, as described in section \ref{sec:ff_theory}.
 The linear fit of the variance gives a diffusion coefficient of $D_{\rm FF,x}$= (1.89 $\pm$ 0.04)$\cdot 10^{9}$ \AA $\mathrm{^{2}/s} $, $D_{\rm FF,y}$= (1.87 $\pm$ 0.04)$\cdot 10^{9}$ \AA $\mathrm{^{2}/s} $, $D_{\rm FF,z}$= (1.89 $\pm$ 0.03)$\cdot 10^{9}$ \AA $\mathrm{^{2}/s} $. This values are in agreement with the one calculated via equation \ref{eq:diff_teo}, being equal to $D_{\rm FF}= 1.86 \cdot 10^{9}$ \AA $\mathrm{^{2}/s} $. This is obtained by summing the contribution of the neighbours till $r_j\approx 80$ \AA\, corresponding to the 8th coordination shell, as already found in \cite{Vittadello2021}. Note that limiting the summation in eq. \ref{eq:diff_teo} to the first neighbors would have seriously underestimated this result.
 
 The temperature dependence of the diffusion coefficient is ruled by the term \ref{eq:MH_trap_prefac} which can be factored out from the summation in eq. \ref{eq:diff_teo}. Thus for free polarons the diffusion is thermally activated with a characteristic energy $U_{FF} = 0.27$ eV given by eq. \ref{eq:hopping_barr}.

\begin{figure}
\centering
\includegraphics{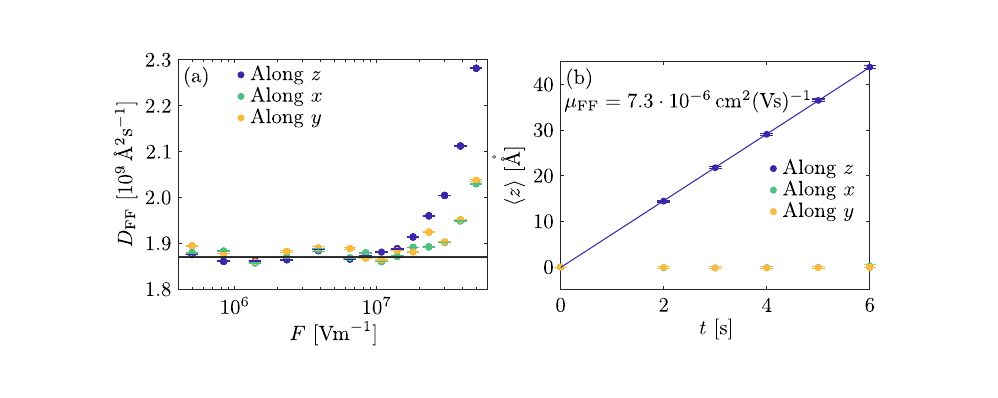}
\caption{(\textbf{a}) Diffusion coefficient in function of the electric field along x, y and z. Horizontal black line corresponds to the value computed with F=0. (\textbf{b}) Mean value of the distribution of the final position in function of the time for the case $\mathrm{F=1 \cdot 10^{7} }$\,V/m. }
\label{fig:free_EF}
\end{figure} 

Let us now consider the effect of an electric field on the free polaron PDF. In fig. \ref{fig:free_EF} (b) it is shown the motion of the polaron distribution as a function of time for a given applied field. As expected the first moment of the polaron distribution is linearly increasing with time, which once again confirms that free polarons obey normal diffusion laws. The slope corresponds to a free polaron mobility at room temperature in stoichiometric LN equal to $\mu = (7.30 \pm 0.3) \cdot 10^{10} $ \AA$^2$/(Vs).  
This result is in agreement with the one expected from the Einstein's relation as long as the applied field is below a value of  $\mathrm{F=1 \cdot 10^{7} }$\,V/m, see fig. \ref{fig:free_EF} (a). For higher field values the variance of the distribution starts to increase because the condition $e\bf{r \cdot F}$ $\ll k_\mathrm{B} T$ is no longer satisfied \cite{Nenashev2010}. 
It is interesting to note that experimentally measured internal fields in cLN at room temperature are generally below this value \cite{Peithmann1999,Vittadello2018}, so that Einstein's relation can indeed be assumed valid for free polaron transport in realistic conditions.

\subsection{Bound polarons hopping}
\label{sec:PP_results}

\begin{figure}
\centering
\includegraphics{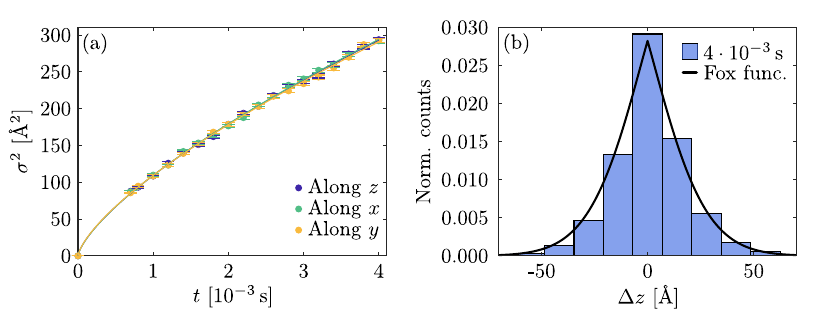}
\caption{(\textbf{a}) Variance of the final electron position as a function of the time for a case of a pure BB jump mechanism at $\mathrm{T=300\,K}$  (\textbf{b}) Probability distribution function of the polaron position at a fixed time at $\mathrm{T=300\,K}$. The histogram is described analytically by the Fox function. }
\label{fig:bound}
\end{figure} 

 Fig. \ref{fig:bound}a) shows the variance of the distribution of the final polaron position along $x$, $y$, and $z$ as a function of time in a case of a pure BB jump mechanism for which free polarons are forbidden in the simulation. The relationship between these two quantities is not linear anymore, confirming that bound polarons are anomalously diffusing on the lattice \cite{Metzler2000}. The simulation results are in agreement with Eq.  \ref{eq:sigma_bound} with $\alpha = 0.72 \pm 0.01 $ and $D_{\rm BB} = (7.1 \pm 0.4) \cdot 10^{3} \, \AA^{2}/\mathrm{s}^{\alpha} $. The value of $\alpha$ reveals that the bound polarons are \textit{sub-diffusing}. The actual value of this parameter is linked to the antisite concentration in the crystal and on the $a_{BB}$ value but not on temperature. 
Also in this case, the thermal dependence of the diffusion coefficient can be factored out in the summation of eq. $\ref{eq:MH_trap_prefac}$. Thus, for pure bound polaron transport, the diffusion coefficient is thermally activated with energy $U_{\rm BB}$ (see eq. \ref{eq:hopping_barr}). Fig. \ref{fig:bound}b) shows the simulated PDF for the bound polarons population at a fixed time. The black line is the the Fox' function reported in equation \ref{eq:Fox_expansion}  with the same $\alpha$ and $D_{\rm BB}$  as reported above, which is in good agreement with simulation results, as expected \cite{Metzler2000}.

\begin{figure}
\centering
\includegraphics{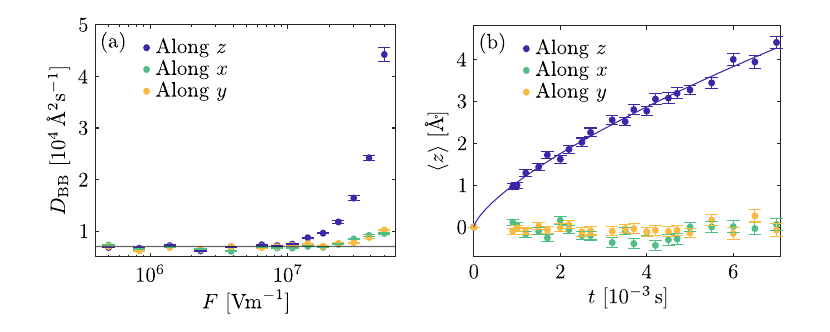}
\caption{(\textbf{a}) Diffusion coefficient in function of the electric field along x, y and z. Horizontal line corresponds to the value computed with F=0. (\textbf{b}) Mean value of the distribution of the final position in function of the time for the case $\mathrm{F=5 \cdot 10^{6} }$\,V/m. }
\label{fig:bound_EF}
\end{figure} 

The effect of an electric field on a polaronic system composed of only bound polarons is reported in Fig. \ref{fig:bound_EF} (b). The PDF average is again changing with time under the effect of the field, but this time with a sub-linear behaviour as expected for anomalous diffusion.  
From the fit with equation \ref{mean_value_bound}, the anomalous mobility and the anomalous coefficient are equal to $\mu_{\alpha} = ( 2.6 \pm 0.4) \cdot 10^{5} \AA^{2}/\mathrm{V}\mathrm{s}^{\alpha}$ and $\alpha=0.71 \pm 0.04$, in agreement with the Einstein's relation.
Also in this case, when the electric field is too strong, the above-mentioned behaviors are no longer observed.
Fig. \ref{fig:bound_EF}a) shows the diffusion coefficient computed in function of the electric field at room temperature. The horizontal line shows the values of the diffusion coefficient computed in the situation $\mathrm{F=0}$ for comparison. Again, above  $\mathrm{F \approx 1 \cdot 10^7\,V/m}$ the variance of the PDF begins to be field-dependent, with a steep increase. 

\subsection{Congruent LN}

In this section, the real-life case of a cLN crystal is considered. As detailed in the Introduction, this material is characterized by a high concentration of intrinsic defects, namely Li vacancies ($\mathrm{V_{Li}}$) and Nb antisites ($\mathrm{Nb_{Li}}$). In this case polarons do not only hop among alike sites, but conversion processes where a free polaron is captured by an antisite defect becoming a bound one and vice-versa are present, mixing up the transport processes described in previous sections. The relative weight of these processes depends not only on the sample composition, but also on temperature \cite{Vittadello2018} as the activation energies for the various processes are different, as per eq. \ref{eq:hopping_barr}.  
Two situations may be considered depending whether the initial delta-like polaron distribution is assumed to be in the free polaron state or in the bound one. This is very much dependent on the experimental conditions and on the time - scale of the experiment one is looking at. In the following both situations are simulated and discussed focusing the attention to the case of a congruent LN crystal with an antisite concentration $[\mathrm{Nb_{Li}}]=1.9 \cdot10^{20}\, \mathrm{cm}^{-3}$ at room temperature. 

\begin{figure}
\centering
\includegraphics{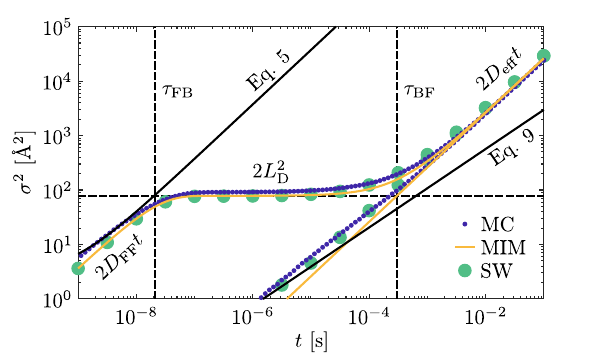}
\caption{Variance of the polaron PDF as a function of the time in a congruent LN crystal at $\mathrm{T=300\,K}$ (blue dots). Theoretical variance according to the switching diffusion model (SW, green dots). Theoretical variance using the MIM model, eq. \ref{eq:doerris_msd} (yellow lines). }
\label{fig:congruent}
\end{figure}

Fig. \ref{fig:congruent} shows the variance of the polaron PDF along the z direction in function of the time for the case that either F or B is the starting polaron of the hopping. The small blue dots are the results obtained from the MC simulations.
If polarons begin their journey as F-polarons (upper curve), three distinct transport phases are visible. Initially, the variance increases almost linearly in the double-logarithmic representation, indicating a power law behaviour. This initial transport phase continues until around 10$^{-7}$\,s when a static phase follows. The transport is then frozen for almost two decades of time, from around $10^{-7}$\,s until $10^{-5}$\,s when the diffusion starts again. In this third phase, the variance evidences again a power-law behaviour. 
If polarons start as B-polaron (lower blue dots), the PDF basically does not change until $10^{-6}$\,s because in such a short time frame the possibility that polarons may escape from the antisites is negligible. From here on, the variance curve shows a linear dependence in the double log plot. This linearity continues until around $10^{-3}$\,s where a slight slope change is visible. 
From here, the evolution of the variance reaches asymptotically the same one observed for a starting F-polaron distribution. 
The physical interpretation of these results is quite straightforward. 
In the case of an initial F-polaron distribution, the particles start to quickly move on the $\mathrm{Nb_{Nb}}$ sublattice until they are gradually trapped by $\mathrm{Nb_{Li}}$  antisites that slow down the diffusion. After this initial stage, polaron motion may occur either by rare $B \rightarrow B$  hopping events or by conversion to free polarons followed by a sequence of hops: $B \rightarrow F \rightarrow F \cdots\rightarrow B$ . 

The green large dots in fig. \ref{fig:congruent} are the numerical solutions for the variance obtained from eq. \ref{eq:switching_differential} assuming the parameters  discussed in the previous sections ($D_{\rm FF}=1.89\cdot10^{9}$ \AA${}^2$/s, $D_{\rm BB}=7.3\cdot10^{3}$ \AA${}^2/s^{\alpha}$, $\alpha=0.71$) for pure F and B transport.  
The switching rates ($k_{\rm FB}=3.76\cdot10^{9}\, \mathrm{s}^{-1}$ , $k_{\rm BF}=2.14\cdot10^{3} \, \mathrm{s}^{-1}$) are computed by comparing the MC simulation results. The characteristic times $\tau_{FB(BF)} = k_{FB(BF)}^{-1}$ mark the positions along the time axis of the knees of the variance curve. 
Finally, the yellow lines in fig. \ref{fig:congruent} are calculated using equations \ref{eq:doerris_msd}, \ref{eq:doerris_msd_startbound} using the same parameters used for the switching diffusion model. As it can be seen, the MIM-model is a good approximation of the observed trends.

\begin{figure}
    \centering
    \includegraphics{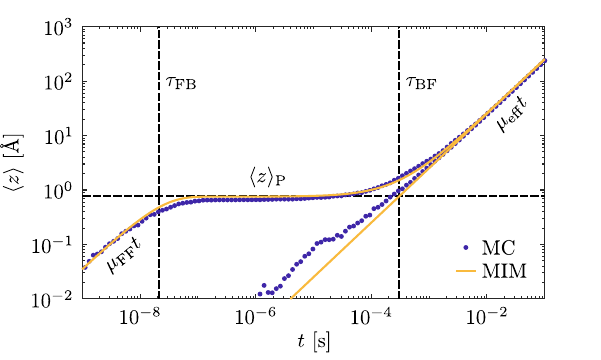}
    \caption{a) Average value of polaron position in function of the time when the starting site is a free polaron (upper blue dots) or bound polaron (lower blue dots) computed for $F = 5\cdot10^6$V/m. Data are superimposed to equation \ref{eq:z_avg_doerris} with the parameters as described in the text.}
    \label{fig:mob_congr}
\end{figure}

Considering now the case of an applied electric field, Fig. \ref{fig:mob_congr} shows the average displacement of the polaron PDF along the polar axis as a function of the time under the effect of an applied bias, here assumed $\mathrm{F=5 \cdot 10^6\,V/m}$. Again, two cases are considered for an initial F-polaron population (upper curve) or for a B-polaron population (lower curve). After an initial swift motion with high mobility equal to $\mu_{FF}$, the polarons remain still under the effect of the field unless sufficient time is elapsed so that polarons can de-trap from antisites and move as free polarons or directly hop on another antisite.
As observed in precedent paragraphs this value is sufficiently low so that Einstein's relationship should hold, which is confirmed by the fact that the variance of the polaron PDF under bias is unchanged with respect to the one considered in fig. \ref{fig:congruent}. 
The simulation results are therefore superposed to eq. \ref{eq:z_avg_doerris} (yellow curve in Fig. \ref{fig:mob_congr}). The parameters are $\mathrm{\tau_{FB}= (2.1 \pm 0.6) \cdot 10^{-8}\,s}$ and $\mathrm{\tau_{BF}= (3 \pm 0.7) \cdot 10^{-4}\,s}$, indicated by the vertical black lines in the figure and in agreement with previous results. 

\section{Discussion}

\begin{table}
    \centering
    \begin{tabular}{c|c|c} \hline 
                                      & $ t\ll \tau_{FB}$ & $ t\gg \tau_{BF}$\\ \hline 
       Stoichiometric LN              & $\mu_{FF} $       &  $\mu_{FF}      $\\ \hline 
       Congruent LN, F starting site  & $\mu_{FF} $       &  $\mu_{eff}     $\\ \hline 
       Congruent LN, B starting site  & $\approx 0$       & $\sim \mu_{eff}      $\\ \hline
    \end{tabular}
    \caption{Limiting values of the polaron mobility for different experimental cases at $T = 300 K$. The parameters are: $\mathrm{\tau_{FB}= (2.1 \pm 0.6) \cdot 10^{-8}\,s}$;  $\mathrm{\tau_{BF}= (3 \pm 0.7) \cdot 10^{-4}\,s}$; $\mu_{FF} = 7.3 \cdot 10^{4} \, \mathrm{cm}^{2} \mathrm{V}^{-1} \mathrm{s}^{-1}$; $\mu_{eff} = 5.1 \, \mathrm{cm}^{2} \mathrm{V}^{-1} \mathrm{s}^{-1}$.  }
    \label{tab:summary}
\end{table}

In the case of defect-free lithium niobate the polaron diffusion and mobility are normal, as expected from theoretical models and verified by Monte-Carlo simulations. The polaron mobility is thermally activated with an energy $U_{FF} = 0.27$ eV and has a value of $\mu_{\rm FF} = 7.3 \cdot 10^{4} \mathrm{cm}^2 / Vs$ at room temperature, corresponding to a diffusion coefficient $D_{\rm FF} = 1.89 \cdot 10^{9} \AA^{2} /s$.
When the more realistic situation of a non-stoichiometric lithium niobate sample is considered, the phenomenology becomes significantly reacher. 
The system can be well described by a switching diffusion model, embodied by eqs. \ref{eq:switching_differential} which can in principle be solved numerically, but for which the rate constants and the anomalous coefficients parameters are difficult to compute analytically. 
Our MC simulations provide for these quantities $\alpha = 0.72 \pm 0.01 $ and $\mathrm{D_{BB}} = (7.1 \pm 0.4) \cdot 10^{3} \, \AA^{2}/\mathrm{s}^{\alpha} $. 
However we showed that for a LN sample in standard conditions, i.e at room temperature and congruent composition (i.e. $\mathrm{Nb_{Li}} = 1.9 \cdot 10^{20} \mathrm{cm}^{-3}$) the influence of direct $B \rightarrow B$ hops is generally small, therefore the system can be described with a good accuracy in terms of a Mobile - Immobile model for which the polaron diffusion and mobility are described by eqs. \ref{eq:doerris_msd}, \ref{eq:doerris_msd_startbound} and \ref{eq:z_avg_doerris}. 
These equations show that the polarons behave differently depending on the experimental situation and on the time scale considered, as detailed below. The main results obtained in the MIM approximation are reported in table \ref{tab:summary}.

\subsection{F starting site}
In experiments where a polaron population is suddenly created e.g. by means of a short laser pulse such as in fs- or TAS spectroscopy, it can be assumed that just after the pulse the majority of the polarons are free. For short times $t\ll \tau_{FB} = \mathrm{k_{FB}}^{-1} $ it is easy to check that eq. \ref{eq:doerris_msd} can be approximated as $\sigma^2 \sim 2D_{FF}t$, so that the system behaves as a defect-free material, with the same mobility. 

\begin{figure}
\centering
\includegraphics[]{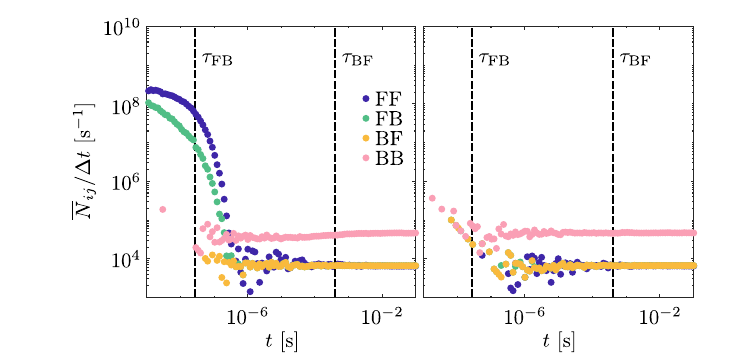}
\caption{Average number of jumps per unit time $\overline{N}_{ij}$, resolved for the different transition types in congruent LN as at T=300K as function of simulation time. a) For the case of polarons starting as F and b) for polarons starting as B. Vertical dashed lines represent $\tau_{\rm FB}$ and $\tau_{\rm BF}$.}
\label{fig:congruent_jumps}
\end{figure}

For $\tau_{\mathrm{FB}}<t<\tau_{\mathrm{BF}}$ F- polarons are almost completely transformed in B- polarons which diffuse much slower than F polarons, so that in this stage the charge diffusion is abruptly decelerated. 
In order to investigate in greater detail the type of conduction in this time frame, we can analyze the relative abundance of the different hop types as retrieved from the MC simulations. In fig. \ref{fig:congruent_jumps} (a) the amount of hops per unit time for the four elementary hopping processes is shown as a function of time. As it can be seen after the initial stage, the number of FF hops drops down to about $10^{4} \,\mathrm{hops}/s$ in the intermediate regime, while $\overline{N}_{FB} = \overline{N}_{BF} $, indicating that the F and B polarons populations are in equilibrium. F polarons do not contribute anymore to the transport, polarons are simply shifting quickly between antisites and regular sites without transport. In addition to those processes, we observe a constant amount of BB hops which is the responsible for the diffusion. Thus, experiments working in this time frame would see the charge carriers diffuse very slowly and anomalously as B polarons with the same parameters described in sec. \ref{sec:PP_results}. 
If this contribution is neglected, we end up with the MIM model description. In this time range eq. \ref{eq:doerris_msd} gives $\sigma^2 \sim 2D_{\rm FF}\tau_{FB} = 2L_{\rm D}^2$ where $L_{\rm D}$ is the diffusion length that can be measured in holographic experiments (see e.g. \cite{Frejlich2006}) and this case equal to $L_{\rm D}\sim 7 \AA$. 
The effect of an applied bias in this time window results in a change of the average polaron position reached when the PDF "freezes" because of the antisites. The plateau value visible in Fig. \ref{fig:mob_congr} can be computed within the MIM model as:
\begin{equation}
     \langle z \rangle_{P}= \frac{e L_{\rm D}^2}{k_BT} F
\end{equation}

Finally, for long times $t \gg \tau_{BF}$, detrapping processes are sufficiently probable so that the chance to have a few polarons contributing to the transport with FF hops is non negligible. Although their number is very limited, the difference in mobility is so large that the FF contribution becomes visible and eventually the dominating one, so that at long times diffusion appears normal again, with an offset deriving from the previous stages.
In the MIM approximation, we obtain from eq. \ref{eq:doerris_msd}:
\begin{equation}
    \sigma^2 \sim 2D_{\rm eff}t + 2L_{\rm D}^2 
    \label{eq:diff_long_time_limit}
\end{equation}
 with $D_{\mathrm{eff}} \approx D_{\rm FF} \frac{\tau_{FB}}{\tau_{BF}} = (1.25 \pm 0.1) \cdot 10^5 $ \AA$^2$/s i.e. about four orders of magnitude smaller than $D_{\rm FF}$. 
In the regime of validity of the Einstein's relation, this corresponds also to the difference in the long-time limit of the polaron mobility between a stoichiometric and a congruent LN crystal at room temperature. 
Indeed, for sufficiently long times, the average displacement under the effect of a field is:
\begin{equation}
    \langle z \rangle \approx \frac{e D_{\rm eff}}{kT} F t +  \langle z \rangle_{P} = \mu_{\rm eff} F t + \langle z \rangle_{P} \sim \mu_{\rm eff} F t
    \label{eq:z_long_time_limit}
\end{equation}
with an effective polaron mobility $\mu_{\rm eff} = 5.1 \, \mathrm{cm}^{2} \mathrm{V}^{-1} \mathrm{s}^{-1} $.

\subsection{B starting site}

Another commonly encountered experimental case is the one of reduced LN samples of congruent composition. If the reduction degree is not too high, charge transport can be attributed to small bound polarons \cite{dhar1990polaronic} which are much more abundant in the lattice than free polarons. Bipolarons are present as well \cite{Schirmer2009}, but they do not appear to contribute significantly to charge transport and in the following will be disregarded. 

At short times ($t \ll \tau_{BF}$), the processes that happen more frequently are BB hops, typically those involving $\mathrm{Nb_{Li}}$ antisites at close distance one to the other. In this regime $D \propto t^{\alpha}$, i.e. the diffusion is anomalous with $\alpha = 0.71$, as visible in Fig. \ref{fig:bound}. While eqs. \ref{eq:switching_differential} may be used for an accurate calculation, the MIM model, which neglects the bound polaron contribution, underestimates the diffusion coefficient in this time frame.
For $t\gg \tau_{BF}$ the role of free polarons is more and more impactful and the situation ends up to the same one considered the previous section, as described by eq. \ref{eq:diff_long_time_limit}, with an effective diffusion coefficient $D_{\rm eff}$. 
As before, the polaron mobility (as shown in Fig. \ref{fig:mob_congr})  follows the behaviour expected from the diffusion coefficient as long as the Einstein's relation holds true.

\section{Conclusions}
The motion of polarons, described by a random hopping process among the sites of a polar crystals is the underlying physical process establishing the electrical properties of the material, which are of paramount importance for several applications. 
In this work the role of defects on polaron diffusion and mobility has been explored theoretically for the specific case of lithium niobate. 
We have shown that the presence of defects requires a significant revision of the theoretical framework needed to describe the polaron motion and obtained realistic estimates of the physical quantities necessary to describe quantitatively the process. 
In particular, the presence of intrinsic defects, as those encountered in the commonly used LN with congruent composition, introduces a time-dependent diffusion and mobility. Depending on the initial sample condition and on the time scale of the experiment one is interested in, several regimes have been evidenced, with the defects affecting the transport behavior to increasing extents. 
Our theoretical models, supported by MC simulations based on the Marcus-Holstein hopping frequency, are able to describe quantitatively the situation under consideration, providing both formal expressions and numerical parameters for the case of congruent LN at room temperature. In particular, the Mobile - Immobile description provides reliable and simple analytical expressions for diffusion and mobility in a time range from $10^{-8}$ s up to the seconds and beyond.

These expressions should replace the ones used in general formulations for diffusion and mobility whenever non-stoichiometric LN is considered. 
Our formalism can be readily extended to other temperatures, provided that the rate constants entering eqs. \ref{eq:switching_differential} are known. Furthermore, our approach may be extended to include additional defects and or dopants with the capability to affect the polaron hopping process, such as Fe traps in the case of Fe-doped lithium niobate or $\mathrm{Ta_{Nb}}$  substitutional defects as in the mixed lithium niobate - tantalate system. 

\section*{References}
\bibliographystyle{unsrt}
\bibliography{manuscript.bib}
\end{document}